\documentclass{aims}
\usepackage{adjustbox}
\usepackage{amsmath}
\usepackage{amsfonts}
\usepackage{amssymb}
\usepackage{graphicx}
\usepackage{color}
\usepackage{url}
\usepackage{mathptmx}
\usepackage{babel}

\usepackage{float}

\usepackage{txfonts}
\def\typeofarticle{Research article}

\def\ppages{213--xxx.}
\def\DOI{10.3934/bioeng.2022015}
\def\Received{22 December 2021}

\def\Accepted{xx June 2022}
\def\Published{xx June 2022}

\usepackage{cite}

\emergencystretch=\maxdimen
\numberwithin{equation}{section}

\makeatletter
\renewcommand{\@biblabel}[1]{#1\hfill \hspace{-0.2cm}}
\makeatother

\begin{document}
	
	\title{Coupled effects of channels and synaptic dynamics  in stochastic modelling of healthy and Parkinson's-disease-affected brains}
	
	\author{%
		Thi Kim Thoa Thieu\affil{1,}\corrauth
		and
		Roderick Melnik\affil{1,2}
	}
	
	\shortauthors{the Author(s)}
	
	\address{%
		\addr{\affilnum{1}}{M3AI Laboratory, MS2Discovery Interdisciplinary Research Institute, Wilfrid Laurier University, Waterloo, Ontario, Canada}
		\addr{\affilnum{2}}{BCAM - Basque Center for Applied Mathematics, Bilbao, Spain}}
	
	\corraddr{Email: tthieu@wlu.ca.}
	
	\begin{abstract}
		Our brain is a complex information processing network in which the nervous system receives information from the environment to quickly react to incoming events or learns from experience to sharp our memory. In the nervous system, the brain states translate collective activities of neurons interconnected via synaptic connections. 
		In this paper, we study coupled effects of channels and synaptic dynamics under the stochastic influence of healthy brain cells with applications to Parkinson's disease (PD). In particular, we investigate the effects of random inputs in a subthalamic nucleus (STN) cell membrane potential model.  
		The STN bursting phenomena and parkinsonian hypokinetic motor symptoms are closely connected, as electrical and chemical maneuvers modulating STN bursts are sufficient to ameliorate or mimic parkinsonian motor deficits. Deep brain stimulation (DBS) of the STN is an important surgical technique used in the treatment to improve PD symptoms. Our numerical results show that the random inputs strongly affect the spiking activities of the STN neuron not only in the case of healthy cells but also in the case of PD cells in the presence of DBS treatment. Specifically, the existence of a random refractory period together with random input current in the system may substantially influence an increased irregularity of spike trains of the output neurons. 
		

	\end{abstract}
	
	
	\maketitle
	
	\section{Introduction}
	
	One of the most common age-associated human neurodegenerative disorders is Parkinson's disease (PD). PD is
	characterized by cardinal motor symptoms such as
	static tremor, bradykinesia, and muscle rigidity. Many different treatments focus on the subthalamic nucleus (STN) to improve such motor symptoms, for instance, ablation surgery of STN or its fiber connections. Deep brain stimulation (DBS) of the STN has recently become an effective therapy of PD  \cite{Yang2014-n}. The DBS is a very impressive method \cite{Shaheen2022} due to the fact that PD,
	characterized by the inadequacy of a chemical substance in
	the brain, can also be successfully treated with 
	passage of only electrical currents without concomitant supply
	of biological or chemical reactions/factors. In general, biological neurons in the brains transmit information by generating spikes and such neurons are connected by synapses that process and store information. 
	To better understand the brain activities, therefore, we need to know how synapses work \cite{Sjostrom2021}. Many models have been proposed to analyze the dynamics of synaptic coupling of human brains in neurodegenerative disorders and therapeutic targets for such diseases (e.g., \cite{Shaheen2021} and references therein). In particular, a model of T-type $\text{Ca}2+$ channels as a new therapeutic target for Parkinson's disease has been proposed in \cite{Yang2014-n}. The authors in \cite{Huang2021} have shown that subthalamic burst discharges play an imperative role in cortico-subcortical information relay,
	and they critically contribute to the pathogenesis of both hypokinetic and hyperkinetic parkinsonian symptoms. The role of the CaV1.3 channels in calcium and iron uptake in the context of
	pharmacological targeting for improving the PD pathology has been discussed in \cite{Boag2021}. A review \cite{Ortner2021} gives the evaluation of the therapeutic potential of  L-type calcium channels (LTCC),  R-type calcium channels (RTCC), and  T-type calcium channels (TTCC)
	inhibition in light of novel preclinical and clinical data and the feasibility of available Ca2+
	channel blockers to modify PD progression. The authors in \cite{Puckerin2018} have considered an engineering selectivity into RGK (Rad, Rem, Rem2, Gem/Kir) GTPase inhibition of
	voltage-dependent calcium channels that is in connection with treatment strategies for diseases including chronic pain
	and Parkinson's disease. Beside the effects of calcium channels on PD, the potassium (K+) channels also play an important role in managing and controlling the PD. There are several results available along this line. In particular, the authors in \cite{Huang2021} have shown  that subthalamic burst discharges
	are dependent on input from the motor
	cortex, causing erroneous re-entrant
	information relays from corticosubthalamic
	to pallido-thalamocortical
	loops and thus parkinsonian tremors.  In \cite{Chen2018}, the authors have summarized the physiological and
	pharmacological effects of three K+ channels as a potential therapeutic target for PD. The effects of pharmacological blockade or activation of K+ channels in the
	progression and treatment of PD have been discussed in \cite{Zhang2020}.

	To get closer to the real scenarios in the application of neuronal models for PD, we should account for the existence of random fluctuations in the system. Specifically, the stochastic inputs arise through sensory fluctuations, brainstem discharges and thermal energy as well as random fluctuations at a microscopic level, such as the Brownian motion of ions.   The stochasticity can arise even from the devices which are used for medical treatments, e.g. devices for injection currents in DBS. The authors in \cite{Powanwe2021} have shown that brain rhythm bursts are enhanced by
	multiplicative noise. The presence of noise in gamma oscillations
	in a model of neuronal networks
	with different reversal potentials has been reported in \cite{Zheng2021}. However, the presence of random factors is not always disadvantageous, such noisy factors can also bring benefits to nervous systems. The noises in the neuronal system are not only a problem for neurons, they can also be a solution in information processing \cite{Faisal2008,Thieu2022pd1}. The detectability of weak signals in nonlinear systems (a phenomenon known as stochastic resonance) can be enhanced by random noise \cite{Groen2019}. The authors in \cite{Groen2019} have also indicated that the phase-based simplification of the
	STN neurons can
	accurately predict responses to temporally complex trains of inputs
	even when the perturbations in timing are large enough to obscure
	the oscillatory nature of the neuron's firing.

	Taking the inspiration from the fields of PD studies together with the effects of natural random factors in biological system dynamics, we develop and investigate a model of coupled effects of channels and synaptic dynamics by using stochastic modelling of healthy brain cells with applications in PD. In particular, we consider a cell membrane potential model in the STN part of the human brain. Our analysis focuses on considering a Langevin stochastic equation in a numerical setting for a cell membrane potential with random inputs. We provide numerical examples and discuss the effects of random inputs on the time evolution of the STN cell membrane potential as well as the spiking activities of the STN neuron.
	Furthermore, we know that the STN bursting phenomenon is one of the main factors that cause parkinsonian hypokinetic
	motor symptoms, whereas. DBS is a surgical technique used in the treatment to improve PD. Our numerical results show that random inputs strongly affect the spiking activities in the STN neuron in the absence and in the presence of DBS.

	\section{Model description}
	
	The second most common neurodegenerative disease after Alzheimer's disease is PD. PD is caused by naturally occurring proteins that fold into the wrong shape and stick together with other proteins, eventually forming thin filament-like structures called amyloid fibrils. Researchers in \cite{Janin2018} have found that calcium influences the way alpha-synuclein proteins interact with synaptic vesicles. In fact, alpha-synuclein is almost like a calcium sensor. In the presence of calcium, alpha-synyclein changes its structure and the way interacts with its environment, which is likely to be very important for its normal function. In  nervous systems,  calcium channels play an important role in the release of neurotransmitters. In particular, when the level of calcium in the nerve cell increases, the alpha-synuclein binds to synaptic vesicles at multiple points causing the vesicles to come together. The normal role of alpha-synuclein is to help the chemical transmission of information across nerve cells. 
	Losing dopaminergic (DA) midbrain neurons within the substantia nigra (SN) can cause prevalent movement disorder. One of the most prevalent disorders is PD.
	In general, a significant increase of the calcium currents in the neuronal system could cause burst discharges in STN. This phenomenon of burst discharges is linked to the loss of DA neurons in the midbrain STN. Therefore, in our model, taking the inspiration from \cite{So2012}, we consider a low-threshold calcium current together with a calcium-activated potassium channel to reduce the effects of calcium currents in the system.


	
	Furthermore, motivated by \cite{So2012,Yang2014-n,Chen2018,Zhang2020}, we consider a modified Hodgkin-Huxley (HH) system modelling a STN cell membrane potential. In particular, we choose first a STN healthy cell, then switch to a PD cell, and study the effects of random inputs on the STN cell membrane potential under synaptic conductance dynamics. 
	
	
	In biological systems of brain networks, instead of physically joined neurons, a spike in the presynaptic cell causes a release of a chemical, or a neurotransmitter. Neurotransmitters are released from synaptic vesicles into a small space between the neurons called the synaptic cleft \cite{Gerstner2014}. In what follows, we will focus on investigating the  chemical synaptic transmission and study how excitation and inhibition affect the patterns in the neurons' spiking output in our HH model.	\clearpage In this section, we consider a HH model of synaptic conductance dynamics. In particular, neurons receive a myriad of excitatory and inhibitory synaptic inputs at dendrites. To better understand the mechanisms of synaptic conductance dynamics, we use the description of Poissonian trains to investigate the dynamics of the random excitatory (E) and inhibitory (I) inputs to a neuron \cite{Dayan2005,Li2019}.

	
	We consider the transmitter-activated ion channels as an explicitly time-dependent conductivity $(g_{\text{syn}}(t))$. The conductance transients can be defined by the following equation (see, e.g., \cite{Dayan2005,Gerstner2014}): 
	
	\begin{align}\label{conductivity}
		\frac{d g_{\text{syn}}(t)}{dt} = -\bar{g}_{\text{syn}}\sum_{k}\delta(t-t_k) - \frac{g_{\text{syn}}(t)}{\tau_{\text{syn}}},
	\end{align}
	where $\bar{g}_{\text{syn}}$ (synaptic weight) denotes the maximum conductance elicited by each incoming spike, while $\tau_{\text{syn}}$ is the synaptic time constant, and $\delta(\cdot)$ is the Dirac delta function. Note that the summation runs over all spikes received by the neuron at time $t_k$.  We have the following formula for converting conductance changes to the current by using Ohm's law:
	
	\begin{align}
		I_{\text{syn}}(t) = g_{\text{syn}}(t)(V(t) - E_{\text{syn}}), 
	\end{align}
	where $V$ is the membrane potential, while $E_{\text{syn}}$ represents the direction of current flow and the excitatory or inhibitory nature of the synapse.

	
	The total synaptic input current $I_{\text{syn}}$ is the combination of both excitatory and inhibitory inputs. Assume that the total excitatory and inhibitory conductances received at time $t$ are $g_E(t)$ and $g_I(t)$, and their corresponding reversal potentials are $E_E$ and $E_I$, respectively. Then, the total synaptic current can be defined by the following equation (see, e.g., \cite{Li2019}): 
	
	\begin{align}
		I_{\text{syn}}(V(t),t) = -g_E(t) (V-E_E) - g_I(t)(V-E_I) = - I_{\text{E}} - I_{\text{I}}. 
	\end{align}
	In \cite{So2012}, the authors have used the quantity $I_{\text{GPe,STN}}$ in the STN model. However, we know that STN-DBS generate
	both excitatory and inhibitory postsynaptic potentials in
	STN neurons \cite{Chiken2016}. In our current consideration, instead of using the current $I_{\text{GPe,STN}}$, we consider the current $I_{\text{STN,DBS}} = - I_{\text{E}} - I_{\text{I}}$.
	%
	%
	Let us define the following synaptic dynamics of the STN cell membrane potential ($V$) described by the following model (based on \cite{So2012})
	\begin{align}\label{main_eq}
		C_m \frac{d}{dt}V(t) &= - I_L - I_{\text{Na}} - I_\text{K} -  I_{\text{T}} - I_\text{Ca} - I_{\text{ahp} } - I_{\text{STN,DBS}} + I_{\text{app}} + I_{\text{dbs}} \quad \text{ if } V(t) \leq V_{\text{th}}, \\
		V(t) &= V_{\text{reset}} \quad \text{ otherwise},
	\end{align}
	where $I_{\text{app}}$ is the external input current, while $C_m$ is the membrane capacitance and $t \in [0, T]$. Additionally, in \eqref{main_eq}, $V_{\text{th}}$ denotes the membrane potential threshold to fire an action potential. \clearpage In this model, we assume that a spike takes place whenever $V(t)$ crosses $V_{\text{th}}$ in the STN membrane potential. In that case, a spike is recorded and $V(t)$ resets to $V_{\text{reset}}$ value. Hence, the reset condition is summarized by $V(t) = V_{\text{reset}}$ if $V(t) \geq V_{\text{th}}$. The quantity $I_{\text{ahp} }$ represents the calcium-activated potassium current for the spike after hyperpolarization in STN.
	
	The concentration of intracellular Ca2+ is governed by the following
	calcium balance equation
	\begin{align}\label{main_eq2}
		\frac{d}{dt}Ca(t) = \varepsilon(I_{\text{Ca} }- I_T - k_\text{Ca}\text{Ca}(t)),
	\end{align}
	where $\varepsilon = 3.75\times10^{-5}$ is a scaling constant, $k_\text{Ca} = 22.5$ (ms$^{-1}$) is a given time constant (see, e.g., \cite{Traub1999,Cornelisse2000}).  
	
	Furthermore, we consider an external random (additive noise) input current as follows:
	$I_{\text{app}} = \mu_{\text{app}}  + \sigma_{\text{app}} \eta(t)$, where $\eta$ is the zero-Gaussian white noise with $\mu_{\text{app}}>0$ and  $\sigma_{\text{app}}>0$. Using the description of such random input current in our system, the first equation \eqref{main_eq} can be considered as the following Langevin stochastic equation (see, e.g., \cite{Roberts2017}):
	
	\begin{align}\label{main_eq02}
		C_m \frac{d}{dt}V(t) &= - I_L - I_{\text{Na}} - I_\text{K} -  I_{\text{T}} - I_\text{Ca}  - I_{\text{ahp}}- I_{\text{E}} - I_{\text{I}} +  I_{\text{dbs}} + \mu_{\text{app}} + \sigma_{\text{app}} \eta(t)
		\quad \text{ if } V(t) \leq V_{\text{th}}. 
	\end{align}
	
	Therefore, the system \eqref{main_eq}--\eqref{main_eq2} ($t \in [0, T]$) can be rewritten as
	\begin{align}
		C_m \frac{d}{dt}V(t) &= - I_L - I_{\text{Na}} - I_\text{K} -  I_{\text{T}} - I_\text{Ca}  - I_{\text{ahp}} - I_{\text{E}} - I_{\text{I}}  +  I_{\text{dbs}} + \mu_{\text{app}} + \sigma_{\text{app}} \eta(t)
		\quad \text{ if } V(t) \leq V_{\text{th}}, \label{eq:1} \\V(t) &= V_{\text{reset}} \quad \text{ otherwise}. \label{eq:2}
	\end{align}
	Furthermore, we consider the following gating variable dynamics (see, e.g., \cite{So2012})
	\begin{align}
		\frac{d}{dt}h(t) &= 0.75\frac{h_{\infty}(V) - h(t)}{\tau_h(V)}, \label{eq:3}\\
		\frac{d}{dt}n(t) &= 0.75\frac{n_{\infty}(V) - n(t)}{\tau_n(V)}, \label{eq:4}\\
		\frac{d}{dt}r(t) &= 0.2\frac{r_{\infty}(V) - r(t)}{\tau_r(V)}, \label{eq:5}\\
		\frac{d}{dt}c(t) &= 0.08\frac{c_{\infty}(V) - c(t)}{\tau_c(V)}, \label{eq:6}\\
		\frac{d}{dt}Ca(t) &= \varepsilon(I_{\text{Ca} }- I_T - k_\text{Ca}\text{Ca}(t)). \label{eq:7}
	\end{align}
	The initial data we use for the system \eqref{eq:1}--\eqref{eq:7} define its initial conditions:
	\clearpage
	\begin{align}
		V(0) &= V_0,  \label{eq:8}\\
		h(0) &= h_\infty(V_0),  \label{eq:9}\\
		n(0) &= n_\infty(V_0),  \label{eq:10}\\
		r(0) &= r_\infty(V_0),  \label{eq:11} \\
		c(0) &= c_\infty(V_0),  \label{eq:12}\\
		Ca(0) &= \frac{a_\infty(V_0)}{a_\infty(V_0) + b_\infty(V_0)},  \label{eq:13}
	\end{align}
	where $h_\infty, n_\infty, r_\infty, c_\infty, a_\infty, b_\infty$ are described as in Table \ref{table:1}. 
	
	In our model \eqref{eq:1}--\eqref{eq:13}, as we mentioned above, we use the simplest input spike train with Poisson process in which the stochastic process of interest provides a suitable approximation to stochastic neuronal firings \cite{Teka2004}. The input spikes will be carried out by the quantity $\sum_{k}\delta(t-t_k)$ in the equation \eqref{conductivity} and the input spikes are given when every input spike arrives independently of other spikes. The process will be described as follows:
	\begin{itemize}
		\item For designing a spike generator of spike train, we define the probability of firing a spike within a short interval  (see, e.g. \cite{Dayan2005}) as $P(1 \text{ spike during } \Delta t) = r_{j}\Delta t$, where $j=e,i$ with $r_e, r_i$ representing the instantaneous excitatory and inhibitory firing rates, respectively.
		\item Then, a Poisson spike train is generated by first subdividing the time interval into a group of short sub-intervals through small time steps $\Delta t$. In our model, we use $\Delta t = 0.1$ (ms).
		\item  We define a random variable $x_{\text{rand}}$ with uniform distribution
		over the range between 0 and 1 at each time step.
		\item  Finally, we compare the random variable $x_{\text{rand}}$ with the probability of firing a spike, which reads:
		
		\begin{align}
			\begin{cases}
				r_j\Delta t > x_{\text{rand}}, \text{ generates a spike},\\
				r_j \Delta t \leq x_{\text{rand}}, \text{ no spike
					is generated}.
			\end{cases}
		\end{align}
	\end{itemize}
	%
	
	By using model \eqref{eq:1}--\eqref{eq:13}, we also investigate the effects of random refractory periods. We consider the random refractory periods $t_{\text{ref}}$ as $t_{\text{ref}} = \mu_{\text{ref}} + \sigma_{\text{ref}}\tilde{ \eta}(t)$, where $\tilde{ \eta}(t) \sim \mathcal{N}(0,1)$ is the standard normal distribution, $\mu_{\text{ref}} > 0$ and $\sigma_{\text{ref}} > 0$.

	In general, the information on stimulating activities in a neuron can be provided by the irregularity of spike trains. The time interval between adjacent spikes is called the inter-spike-interval (ISI). The coefficient of variation (CV) of the ISI in a cell membrane potenial with multiple inputs can bring useful information about the output of a decoded neuron. In what follows, we will demonstrate that when we increase the value of $\sigma_{\text{ref}}$, the irregularity of the spike trains increases (see also \cite{Gallinaro2021}). 
	%
	\clearpage
	The spike irregularity of spike trains can be described via the coefficient of variation of the inter-spike-interval (see, e.g., \cite{Christodoulou2001,Gallinaro2021}) as follows:
	\begin{align}\label{CV}CV_{\text{ISI}} = \frac{\sigma_\text{ISI}}{\mu_\text{ISI}},\end{align}
	where $\sigma_\text{ISI}$ is the standard deviation and $\mu_\text{ISI}$ is the mean of the ISI of an individual neuron. 
	
	In the next section, let us consider the output firing rate as a function of Gaussian white noise mean or direct current value, namely, the input-output transfer function of the neuron.

	
	
	In our model, we choose the parameter set as in the following Table \ref{table:1}:
	\begin{table}[H]
		\centering
		\caption{Steady-state functions for channel gating variables and time constants for the different ion
			channels (see, e.g., \cite{So2012}).}
		\label{table:1}
		\begin{adjustbox}{width=1\textwidth}
			\begin{tabular}{l l l l l} 
				\hline
				Current  & Gating variables & Gating variables & Parameters \\ [0.5ex] 
				\hline
				$I_{\text{L}} = g_L(v- E_{\text{L}})$ &  &  & $g_L = 2.25$ (nS)\\   &  &  & $E_\text{L} = -60$ (mV)\\ 
				
				$I_\text{Na} = g_\text{Na}m_{\infty}^3(V)h(V)(V- E_{\text{Na}})$ & $m_{\infty} (V) = 1/(1+ \exp(-\frac{V+30}{15}))$ & $h_\infty(V) = 1/(1+ \exp(-\frac{V+39}{3.1})$ & $g_\text{Na} = 37$ \\
				&   &$\tau_h(V) =1+ 500/(1+ \exp(-\frac{V+57}{-3})$  & $E_\text{Na} = 55$ (mV)\\ 
				
				$I_{\text{K}} = g_\text{K}n^4(V)(V - E_\text{K})$ & $n_{\infty} (V) = 1/(1+ \exp(-\frac{V+32}{8}))$ &  & $g_\text{K} = 45$ (nS)  \\&  $\tau_n(V) =1+ 100/(1+ \exp(-\frac{V+80}{-26})$&  & $E_\text{K} = -80$ (mV)\\ 
				
				$I_\text{T} = g_\text{T}a^3_\infty(V)b_\infty^2(r)r(V)(V - E_\text{T})$ & $a_{\infty} (V) = 1/(1+ \exp(-\frac{V+63}{7.8}))$ & $r_\infty(V) = 1/(1+ \exp(\frac{V+67}{2})$ & $g_\text{T} = 0.5$ (nS)\\
				& $b_{\infty} (V) = 1/(1+ \exp(-\frac{V-0.4}{0.1}))$ & $\tau_r(V) =7.1+ 17.5/(1+ \exp(-\frac{V+68}{-2.2})$ & $E_\text{T} = 0$ (mV)\\& $ - 1/(1+ \exp(4))$  &  & \\ 
				
				$I_\text{Ca} = g_\text{Ca}c^2(V)(V- E_{\text{Ca}})$ & $c_{\infty} (V) = 1/(1+ \exp(-\frac{V+20}{8}))$ &  & $g_\text{Ca} = 2$ (nS) \\ &$\tau_c(V) =1+ 10/(1+ \exp(\frac{V+80}{26})$  &  & $E_\text{Ca} = 140$ (mV)\\ 
				
				$I_\text{ahp} = g_\text{ahp}(V- E_\text{ahp})(\frac{\text{Ca}}{\text{Ca} + 15})$ &  &  & $g_\text{ahp} = 20$ (nS) \\ &  &  & $E_\text{ahp} = -80$ (mV)\\ [1ex] 
				
				$I_{\text{dbs}} = 5+5\sin(2\pi t)$ (pA)&  &  & \\ [1ex] 
				\hline
			\end{tabular}
		\end{adjustbox}
		
	\end{table}
	
	Since these parameters have also been used in \cite{So2012} for STN cell membrane potential experiments, we take them for our model validation. Moreover, in our consideration, we use not only the parameters from Table \ref{table:1}, but also the following parameters: $V_{\text{th}} = -55$ (mV), $V_{\text{reset}} = -70$ (mV), $V_0 = -65$ (mV), $\Delta t = 0.1$, $C_m = 10$ (nF), $\tau_{E} =2$ (ms), $\tau_{I} = 5$ (ms), $\bar{g}_E = 1.5$ (nS), $\bar{g}_I = 0.5$ (nS), $r_e = 10$, $r_i = 10$, $n_E = 20$ spike trains, $n_I = 80$ spike trains. Here, $n_E$ and $n_I$ represent the number of excitatory and inhibitory presynaptic spike trains, respectively. 
	
	
	Mathematically, the developed model \eqref{eq:1}--\eqref{eq:13} is an evolutionary system that combines stochastic differential equations and ordinary differential equations (SDEs-ODEs), where the stochastic membrane potential equation is coupled to the activation and inactivation ion channels equations, as well as to the calcium-activated potassium current equation. This system can be considered as a modified HH system. 
	
	Models proposed in \cite{So2012} represent the responses of STN neurons to the depolarization of current injection
	with repetitive firing and exhibit rebound bursts with more rapid de-activation of T-type calcium currents $I_\text{T}$. However, in their models, they consider a modified version of a basal Ganglia-Thalamic network model with a special focus on the dynamics of membrane potentials in a deterministic case. In real-world applications, the stochastic factors become important in capturing the effects of ion channels. Taking the inspiration from the studies in \cite{Breakspear2017-rv,Roberts2017}, our focus in the remainder of this paper will be on the effects of random inputs on a STN cell membrane potential under synaptic dynamics with applications in PD. In general, the external current controls the firing mode in neuronal systems. A wrong thalamic transmission could lead to errors such as misses, bursts and/or spurious events. Moreover, the contribution of random factors could reduce the response of the neuron to each stimulus in the STN. In the next section, we will be analyzing the effects of Gaussian white noise input current and random refractory periods on the spiking activities in a STN cell membrane potential in the absence and in the presence of DBS.


	\section{Numerical results}

	In this section, we take a single STN neuron and study how the neuron behaves under random inputs and when it is bombarded with both excitatory and inhibitory spike trains.
	The numerical results reported in this section are obtained by using a discrete-time integration based on the Euler method implemented in Python. 
	
	In particular, we use the coupled SDEs-ODEs system \eqref{eq:1}--\eqref{eq:13} that describes the dynamics of the STN membrane potential. As we have mentioned in the previous section, we will focus on the effects of Gaussian white noise input current together with the random refractory periods on the STN cell membrane potential. 
	
\begin{figure}[h!]
		\centering
		\includegraphics[width=0.8\textwidth]{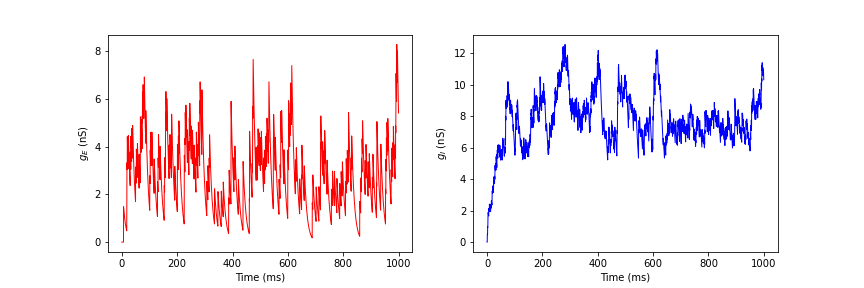} 
		\caption{ Left: Excitatory conductances profile corresponding to the dynamics \eqref{conductivity}. Right: Inhibitory conductances profile corresponding to the dynamics \eqref{conductivity}.}
		\label{fig:0-0}
	\end{figure}

\begin{figure}[h!]
		\centering
		\includegraphics[width=0.8\textwidth]{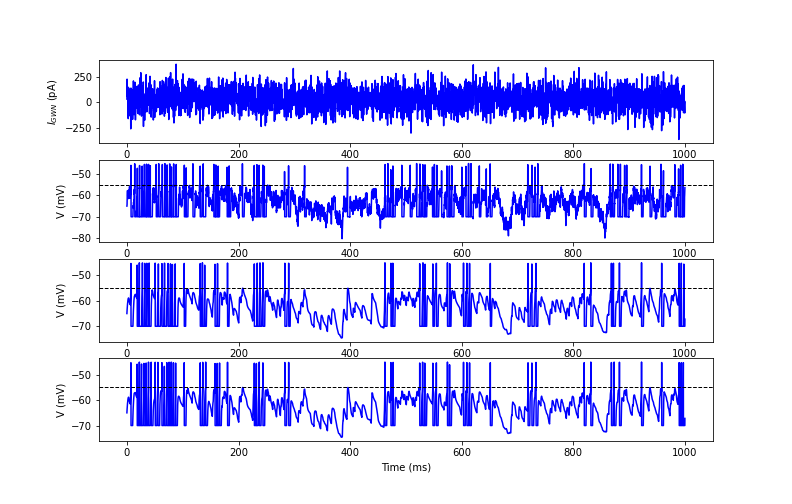}
		\caption{First row: Gaussian white noise current profile. Second row: Time evolution of membrane potential $V(t)$ with additive noise current $I_\text{app} = 33 + \eta(t)$ (pA) and random refractory period $t_\text{ref} = 3 + \sigma_{\text{ref}}\tilde{ \eta}(t)$ (ms). Third row: Time evolution of membrane potential $V(t)$ with direct input current $I_\text{app} = 33$ (pA) and with random refractory period $t_\text{ref} = 3 +  \sigma_{\text{ref}}\tilde{ \eta}(t)$ (ms). Fourth row: Time evolution of membrane potential $V(t)$ with direct input current $I_\text{app} = 33$ (pA) and with direct refractory period $t_\text{ref} = 3$ (ms). Parameters: $ \sigma_{\text{ref}}=0.5$, $I_{\text{dbs}} = 0$ (pA). The dash line represents the spike threshold $V_{\text{th}} = -55$ (mV).}
		\label{fig:0-1}
	\end{figure}

\begin{figure}[h!]
		\centering
		\includegraphics[width=0.8\textwidth]{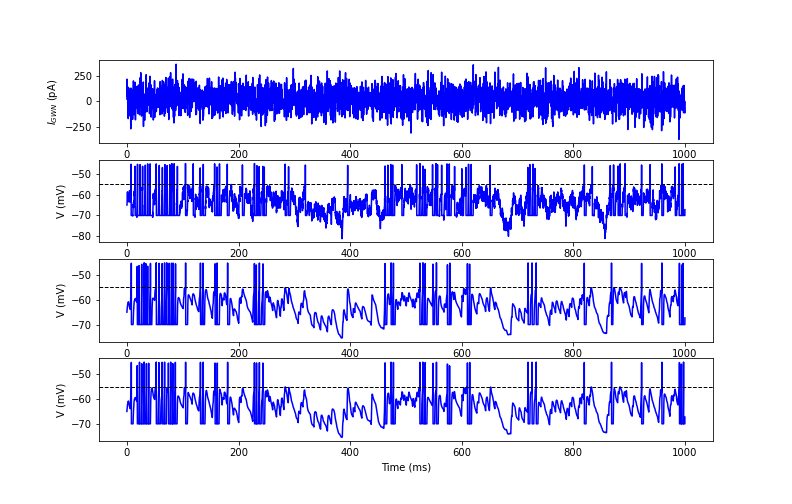}
		\caption{First row: Gaussian white noise current profile. Second row: Time evolution of membrane potential $V(t)$ with additive noise current $I_\text{app} = 23 + \eta(t)$ (pA) and random refractory period $t_\text{ref} = 3 + \tilde{ \eta}(t)$ (ms). Third row: Time evolution of membrane potential $V(t)$ with direct input current $I_\text{app} = 23$ (pA) and with random refractory period $t_\text{ref} = 3 +\tilde{ \eta}(t)$ (ms). Fourth row: Time evolution of membrane potential $V(t)$ with direct input current $I_\text{app} = 23$ (pA) and with direct refractory period $t_\text{ref} = 3$ (ms). Parameters: $ \sigma_{\text{ref}}=0.5$, $I_{\text{dbs}} = 0$ (pA). The dash line represents the spike threshold $V_{\text{th}} = -55$ (mV).}
		\label{fig:0-2}
	\end{figure}

\begin{figure}[h!]
		\centering
		\includegraphics[width=0.8\textwidth]{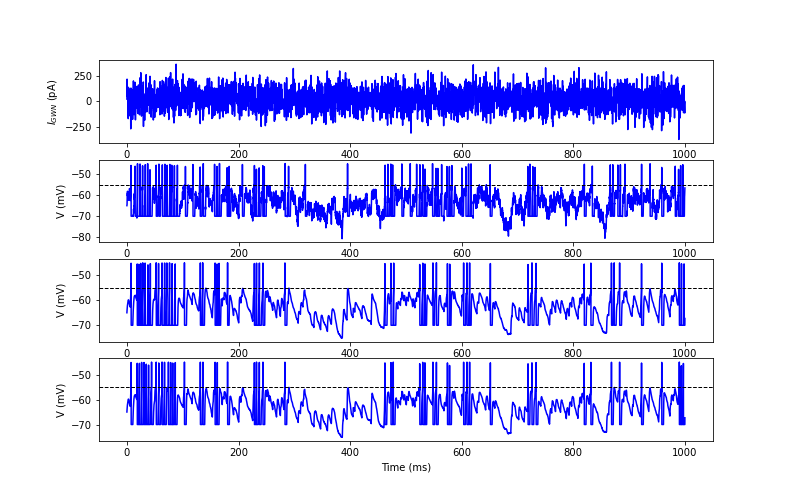}
		\caption{ First row: Gaussian white noise current profile. Second row: Time evolution of membrane potential $V(t)$ with additive noise current $I_\text{app} = 23 + \eta(t)$ (pA) and random refractory period $t_\text{ref} = 3 + \tilde{ \eta}(t)$ (ms). Third row: Time evolution of membrane potential $V(t)$ with direct input current $I_\text{app} = 23$ (pA) and with random refractory period $t_\text{ref} = 3 + \tilde{ \eta}(t)$ (ms). Fourth row: Time evolution of membrane potential $V(t)$ with direct input current $I_\text{app} = 23$ (pA) and with direct refractory period $t_\text{ref} = 3$ (ms). Parameters: $ \sigma_{\text{ref}}=0.5$, $I_{\text{dbs}} = 5+5\sin(2\pi t)$ (pA). The dash line represents the spike threshold $V_{\text{th}} = -55$ (mV).}
		\label{fig:0-3}
	\end{figure}

\begin{figure}[h!]
		\centering
		\includegraphics[width=0.8\textwidth]{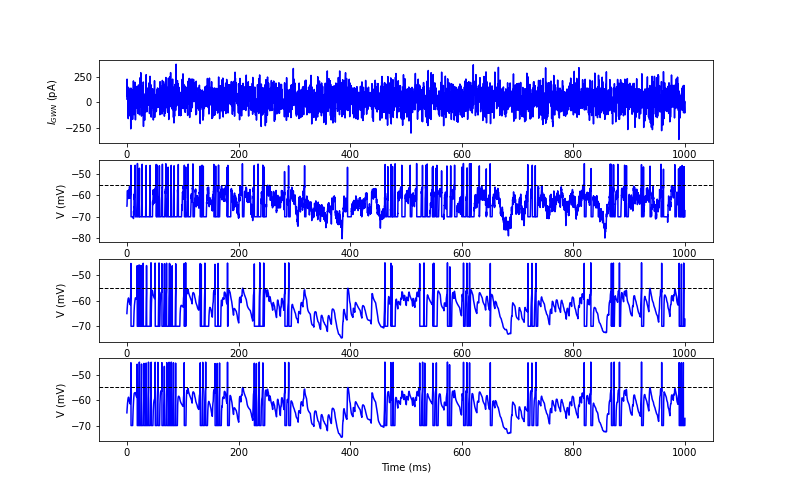}
		\caption{ First row: Gaussian white noise current profile. Second row: Time evolution of membrane potential $V(t)$ with additive noise current $I_\text{app} = 33 + \eta(t)$ (pA) and random refractory period $t_\text{ref} = 3 + \sigma_{\text{ref}}\tilde{ \eta}(t)$ (ms). Third row: Time evolution of membrane potential $V(t)$ with direct input current $I_\text{app} = 33$ (pA) and with random refractory period $t_\text{ref} = 3 +  \sigma_{\text{ref}}\tilde{ \eta}(t)$ (ms). Fourth row: Time evolution of membrane potential $V(t)$ with direct input current $I_\text{app} = 33$ (pA) and with direct refractory period $t_\text{ref} = 3$ (ms). Parameters: $ \sigma_{\text{ref}}=2$, $I_{\text{dbs}} = 0$ (pA). The dash line represents the spike threshold $V_{\text{th}} = -55$ (mV).}
		\label{fig:0-4}
	\end{figure}

\begin{figure}[h!]
		\centering
		\includegraphics[width=0.8\textwidth]{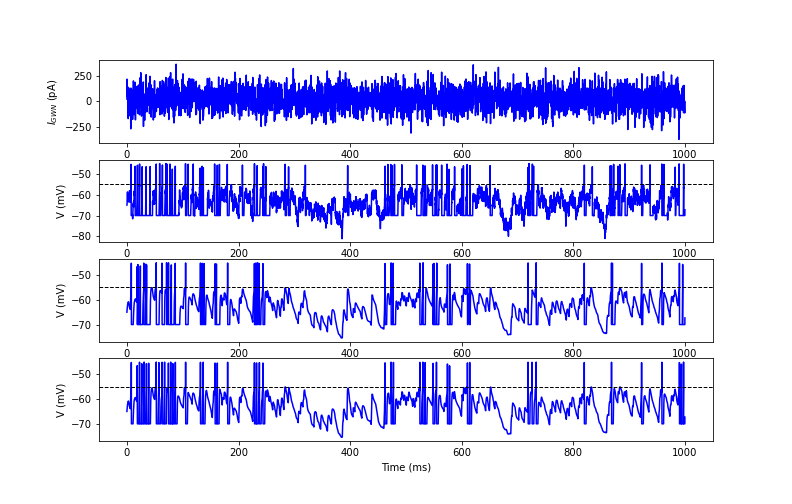}
		\caption{ First row: Gaussian white noise current profile. Second row: Time evolution of membrane potential $V(t)$ with additive noise current $I_\text{app} = 23 + \eta(t)$ (pA) and random refractory period $t_\text{ref} = 3 + \tilde{ \eta}(t)$ (ms). Third row: Time evolution of membrane potential $V(t)$ with direct input current $I_\text{app} = 23$ (pA) and with random refractory period $t_\text{ref} = 3 + \tilde{ \eta}(t)$ (ms). Fourth row: Time evolution of membrane potential $V(t)$ with direct input current $I_\text{app} = 23$ (pA) and with direct refractory period $t_\text{ref} = 3$ (ms). Parameters: $ \sigma_{\text{ref}}=2$, $I_{\text{dbs}} = 0$ (pA).The dash line represents the spike threshold $V_{\text{th}} = -55$ (mV). }
		\label{fig:0-5}
	\end{figure}

\begin{figure}[h!]
		\centering
		\includegraphics[width=0.8\textwidth]{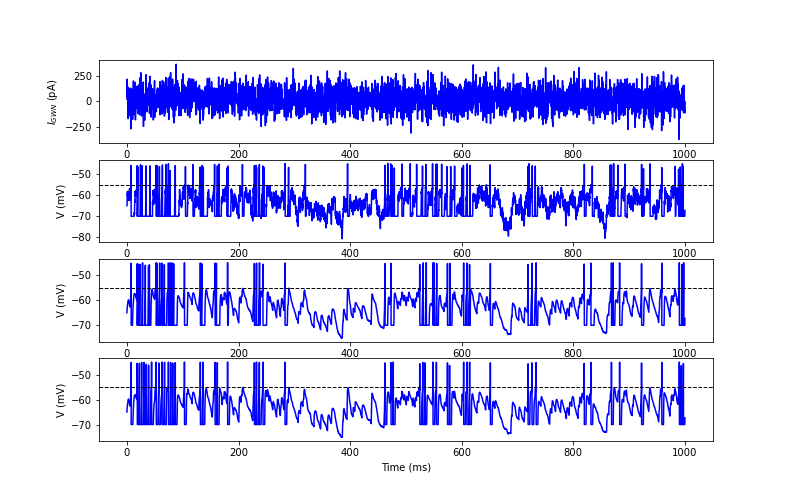}
		\caption{  First row: Gaussian white noise current profile. Second row: Time evolution of membrane potential $V(t)$ with additive noise current $I_\text{app} = 23 + \eta(t)$ (pA) and random refractory period $t_\text{ref} = 3 + \tilde{ \eta}(t)$ (ms). Third row: Time evolution of membrane potential $V(t)$ with direct input current $I_\text{app} = 23$ (pA) and with random refractory period $t_\text{ref} = 3 + \tilde{ \eta}(t)$ (ms). Fourth row: Time evolution of membrane potential $V(t)$ with direct input current $I_\text{app} = 23$ (pA) and with direct refractory period $t_\text{ref} = 3$ (ms). Parameters: $ \sigma_{\text{ref}}=2$, $I_{\text{dbs}} = 2+\sin(2\pi t)$ (pA). The dash line represents the spike threshold $V_{\text{th}} = -55$ (mV).}
		\label{fig:0-6}
	\end{figure}

\begin{figure}[h!]
		\centering
		\includegraphics[width=0.8\textwidth]{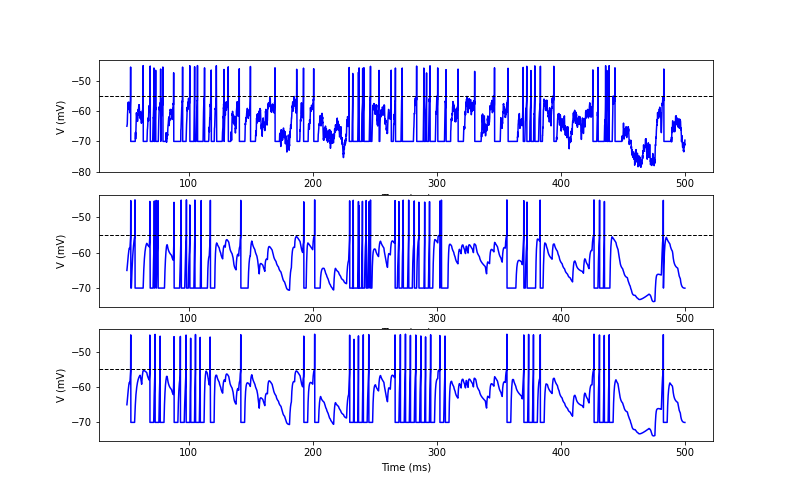}
		\caption{ First row: Gaussian white noise current profile. Second row: Time evolution of membrane potential $V(t)$ with additive noise current $I_\text{app} = 23 + \eta(t)$ (pA) and random refractory period $t_\text{ref} = 3 + \tilde{ \eta}(t)$ (ms). Third row: Time evolution of membrane potential $V(t)$ with direct input current $I_\text{app} = 23$ (pA) and with random refractory period $t_\text{ref} = 3 + \tilde{ \eta}(t)$ (ms). Fourth row: Time evolution of membrane potential $V(t)$ with direct input current $I_\text{app} = 33$ (pA) and with direct refractory period $t_\text{ref} = 3$ (ms). Parameters: $ \sigma_{\text{ref}}=2$, $I_{\text{dbs}} = 0$ (pA). The dash line represents the spike threshold $V_{\text{th}} = -55$ (mV).}
		\label{fig:0-6-1}
	\end{figure}

\begin{figure}[h!]
		\centering
		\includegraphics[width=0.8\textwidth]{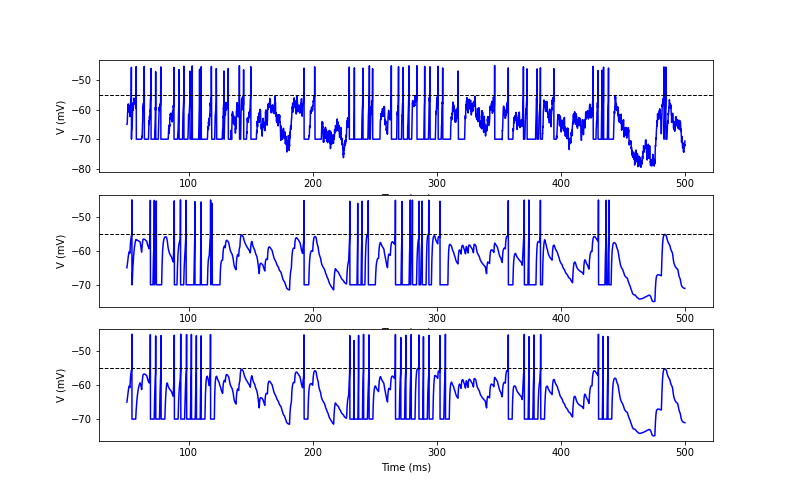}
		\caption{ First row: Gaussian white noise current profile. Second row: Time evolution of membrane potential $V(t)$ with additive noise current $I_\text{app} = 23 + \eta(t)$ (pA) and random refractory period $t_\text{ref} = 3 + \tilde{ \eta}(t)$ (ms). Third row: Time evolution of membrane potential $V(t)$ with direct input current $I_\text{app} = 23$ (pA) and with random refractory period $t_\text{ref} = 3 + \tilde{ \eta}(t)$ (ms). Fourth row: Time evolution of membrane potential $V(t)$ with direct input current $I_\text{app} = 23$ (pA) and with direct refractory period $t_\text{ref} = 3$ (ms). Parameters: $ \sigma_{\text{ref}}=2$, $I_{\text{dbs}} = 0$ (pA).  The dash line represents the spike threshold $V_{\text{th}} = -55$ (mV).}
		\label{fig:0-6-2}
	\end{figure}

\begin{figure}[h!]
		\centering
		\includegraphics[width=0.8\textwidth]{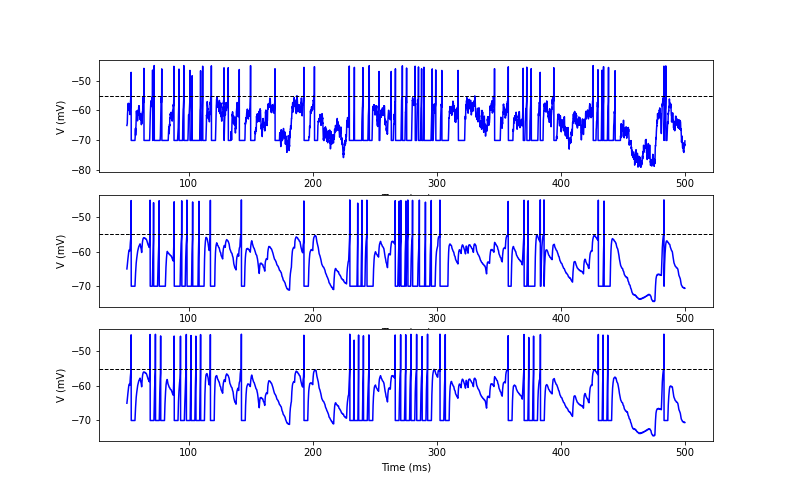}
		\caption{ First row: Gaussian white noise current profile. Second row: Time evolution of membrane potential $V(t)$ with additive noise current $I_\text{app} = 23 + \eta(t)$ (pA) and random refractory period $t_\text{ref} = 3 + \tilde{ \eta}(t)$ (ms). Third row: Time evolution of membrane potential $V(t)$ with direct input current $I_\text{app} = 23$ (pA) and with random refractory period $t_\text{ref} = 3 + \tilde{ \eta}(t)$ (ms). Fourth row: Time evolution of membrane potential $V(t)$ with direct input current $I_\text{app} = 23$ (pA) and with direct refractory period $t_\text{ref} = 3$ (ms). Parameters: $ \sigma_{\text{ref}}=2$, $I_{\text{dbs}} = 2+\sin(2\pi t)$ (pA). The dash line represents the spike threshold $V_{\text{th}} = -55$ (mV).}
		\label{fig:0-6-3}
	\end{figure}

\begin{figure}[h!]
		\centering
		\begin{tabular}{ll}
			\includegraphics[width=0.4\textwidth]{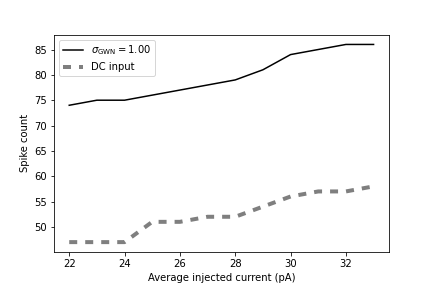} &\includegraphics[width=0.4\textwidth]{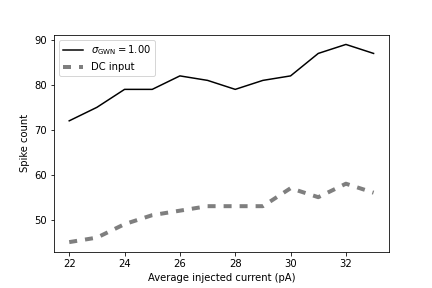} \\
			\includegraphics[width=0.4\textwidth]{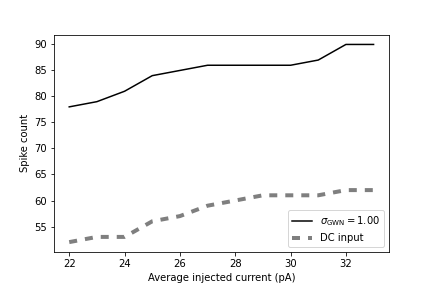} &\includegraphics[width=0.4\textwidth]{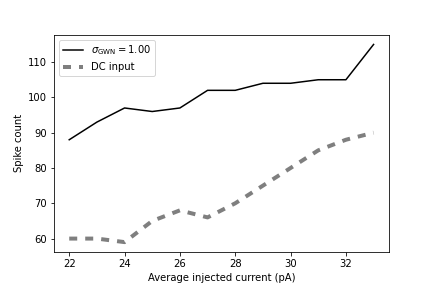}
		\end{tabular}
		\caption{ The input-output transfer function of the neuron with the output firing rate as a function of input mean for the case with additive noise input current ($\sigma_{\text{app}} = 1$).  Top left panel: direct time refractory period $t_{\text{ref}} = 3$ (ms) with $I_{\text{dbs}}=0$ (pA). Top right panel: random refractory period $t_{\text{ref}} = 3 + 2\tilde{ \eta}(t)$ (ms) with $I_{\text{dbs}}=0$. Bottom left panel: direct time refractory period $t_{\text{ref}} = 3$ (ms) with $I_{\text{dbs}}=2+\sin(2\pi t)$ (pA). Bottom right panel: random refractory period $t_{\text{ref}} = 3 + 2\tilde{ \eta}(t)$ (ms) with $I_{\text{dbs}}=5+5\sin(2\pi t)$ (pA).
		}\label{fig:0-7}	
	\end{figure}

\begin{figure}[h!]
		\centering
		\begin{tabular}{ll}
			\includegraphics[width=0.48\textwidth]{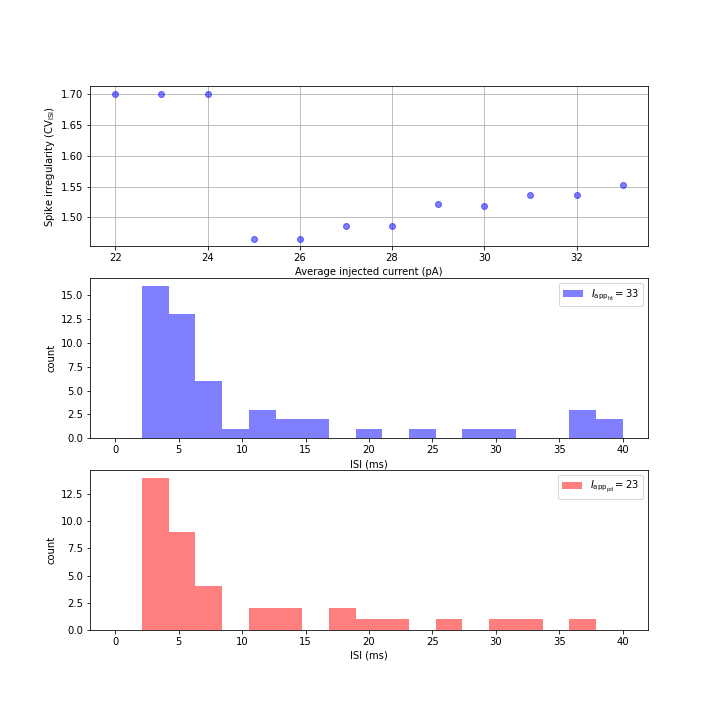}&
			\includegraphics[width=0.48\textwidth]{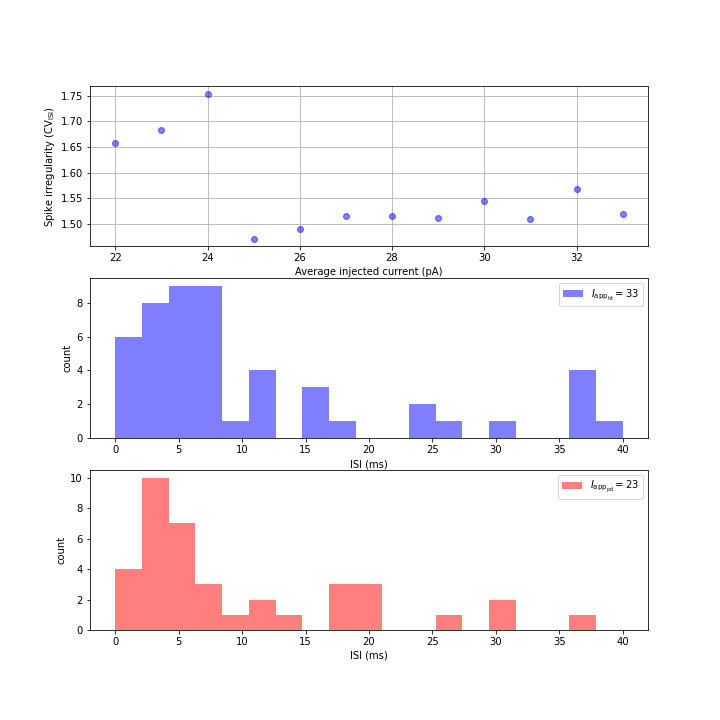}
		\end{tabular}
		\caption{ Spike irregularity profiles in the case with the direct input current and the ISI distribution. First column: direct time refractory period $t_{\text{ref}} = 3$ (ms). Second column: random refractory period $t_{\text{ref}} = 3 +\tilde{ \eta}(t)$ (ms). Middle row: ISI distribution of a healthy cell. Last row: ISI distribution of a PD-affected cell. 
		}\label{fig:0-8-0}	
	\end{figure}

\begin{figure}[h!]
		\centering
		\begin{tabular}{ll}
			\includegraphics[width=0.48\textwidth]{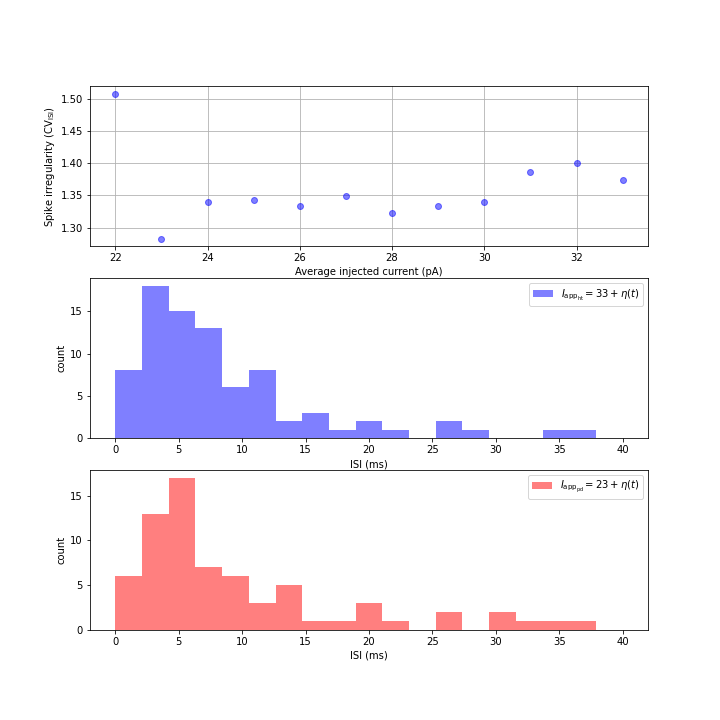}&
			\includegraphics[width=0.48\textwidth]{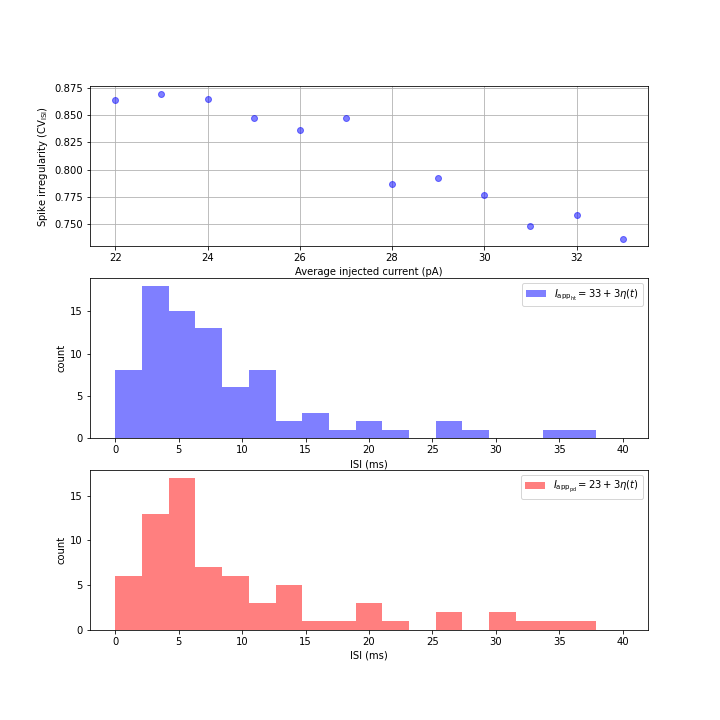} 
		\end{tabular}
		\caption{ Spike irregularity profiles in the case with the additive noise input current and the ISI distribution. First column: random refractory period $t_{\text{ref}} = 3 +2\tilde{ \eta}(t)$ (ms) with $I_\text{app} = I_{\text{average}} + \eta(t)$. Second column: random refractory period $t_{\text{ref}} = 3 +2\tilde{ \eta}(t)$ (ms) with $I_\text{app} = I_{\text{average}} + 3\eta(t)$. Middle row: ISI distribution of a healthy cell in presence of Gaussian white noise $I_\text{app} = 33 + \eta(t)$ (pA). Last row: ISI distribution of a PD-affected cell in presence of Gaussian white noise $I_\text{app} = 23 + \eta(t)$ (pA).  
		}\label{fig:0-8-2}	
	\end{figure}

\begin{figure}[h!]
		\centering
		\begin{tabular}{ll}
			\includegraphics[width=0.48\textwidth]{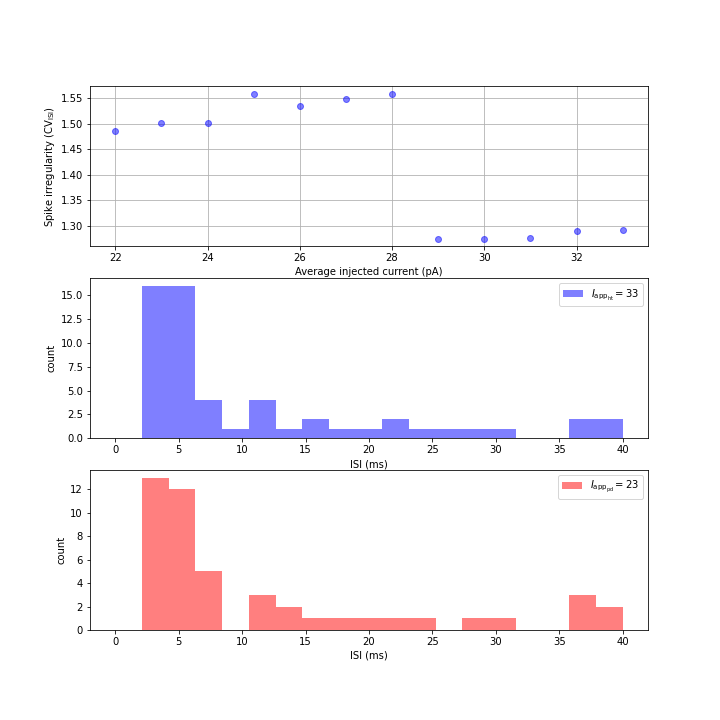}  &
			\includegraphics[width=0.48\textwidth]{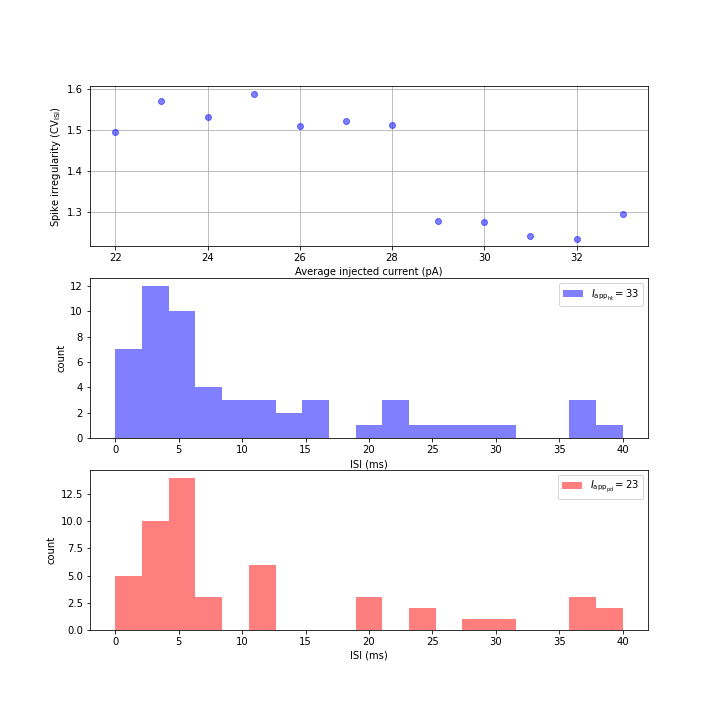}  
		\end{tabular}
		\caption{Spike irregularity profiles in the case with the additive noise input current and the ISI distribution. First column: direct time refractory period $t_{\text{ref}} = 3$ (ms) with $I_{\text{dbs}}=5+5\sin(2\pi t)$ (pA). Second column: random refractory period $t_{\text{ref}} = 3 +\tilde{ \eta}(t)$ (ms) with $I_{\text{dbs}}=5+5\sin(2\pi t)$ (pA). Middle row: ISI distribution of a healthy cell. Last row: ISI distribution of a PD-affected cell. 
		}\label{fig:0-8-1}	
	\end{figure}

\begin{figure}[h!]
		\centering
		\begin{tabular}{ll}
			\includegraphics[width=0.48\textwidth]{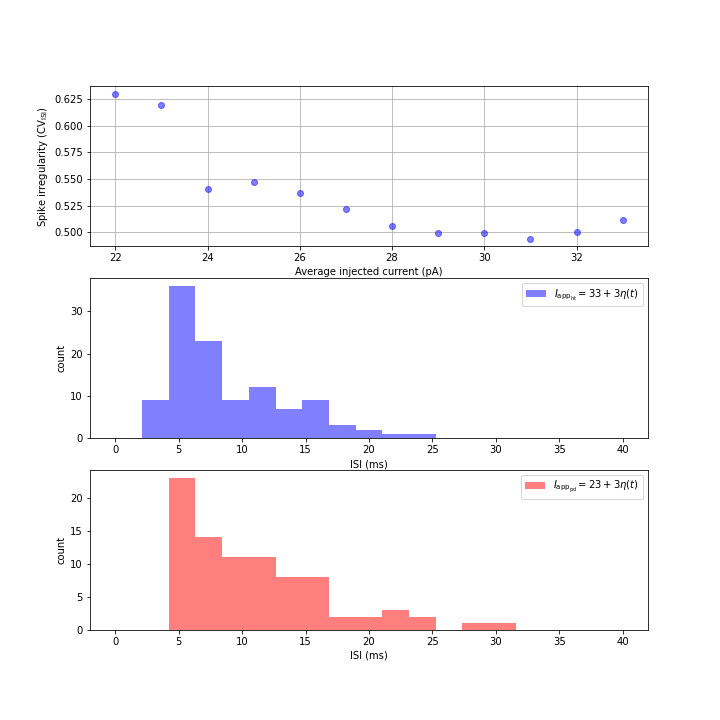} &\includegraphics[width=0.48\textwidth]{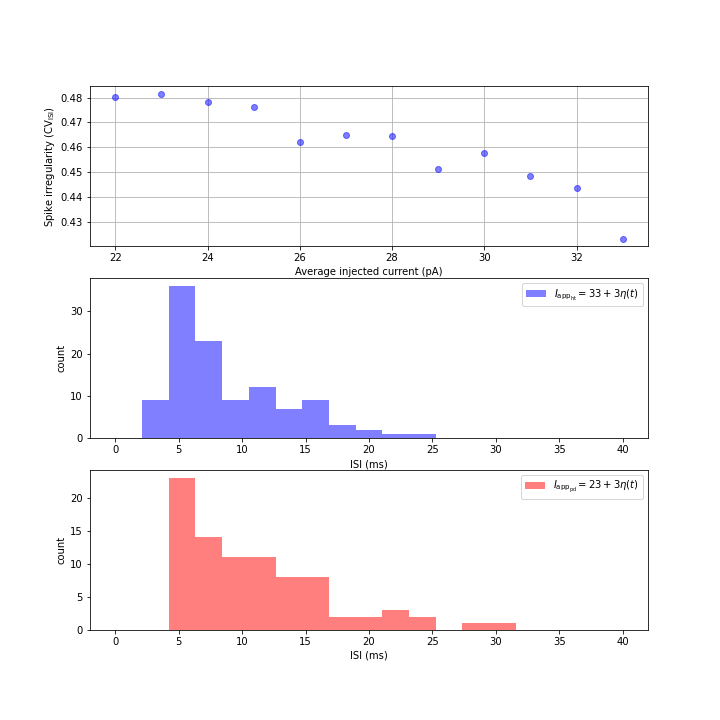} 
		\end{tabular}
		\caption{ Spike irregularity profiles in the case with the additive noise input current and the ISI distribution. First column: direct time refractory period $t_{\text{ref}} = 3$ (ms) with $I_{\text{dbs}}=5+5\sin(2\pi t)$ (pA). Second column: direct time refractory period $t_{\text{ref}} = 3$ (ms) with $I_{\text{dbs}}=5+5\sin(2\pi t)$ (pA). Middle row: ISI distribution of a healthy cell in presence of Gaussian white noise $I_\text{app} = 33 + \eta(t)$ (pA) (left) and $I_\text{app} = 33 + 3\eta(t)$ (pA) (right). Last row: ISI distribution of a PD-affected cell in presence of Gaussian white noise $I_\text{app} = 23 + \eta(t)$ (pA) (left) and $I_\text{app} = 23 + 3\eta(t)$ (pA) (right).  
		}\label{fig:0-8-3}	
	\end{figure}

\begin{figure}[h!]
		\centering
		\begin{tabular}{ll}
			\includegraphics[width=0.48\textwidth]{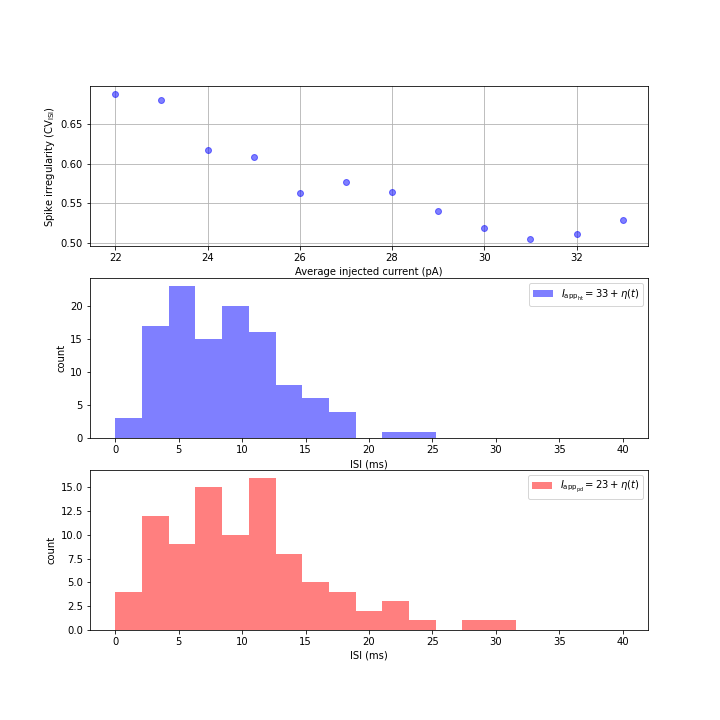}  &
			\includegraphics[width=0.48\textwidth]{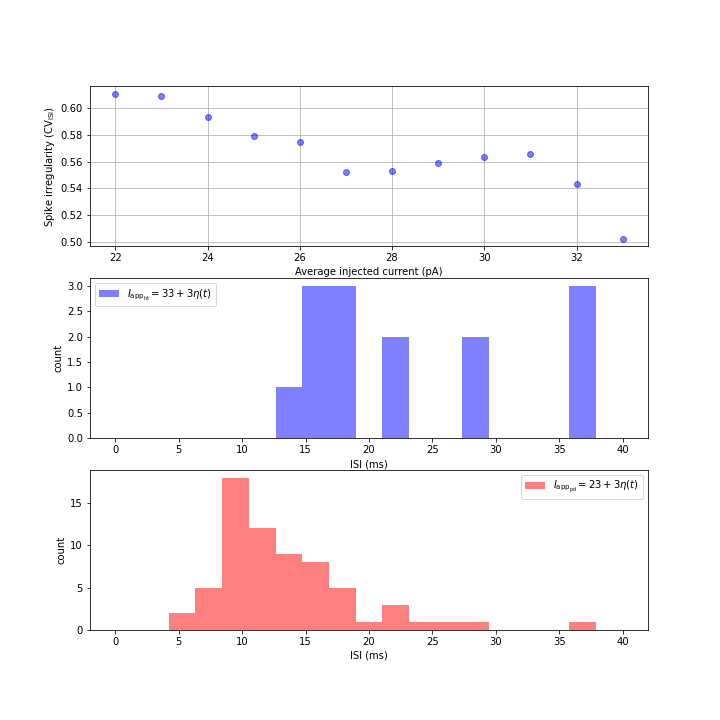} 
		\end{tabular}
		\caption{ Spike irregularity profiles in the case with the additive noise input current and the ISI distribution. First column: random refractory period $t_{\text{ref}} = 3 +2\tilde{ \eta}(t)$ (ms) with $I_{\text{dbs}}=2+\sin(2\pi t)$ (pA). Second column: random refractory period $t_{\text{ref}} = 3 +2\tilde{ \eta}(t)$ (ms) with $I_{\text{dbs}}=2+\sin(2\pi t)$ (pA). Middle row: ISI distribution of a healthy cell in presence of Gaussian white noise $I_\text{app} = 33 + \eta(t)$ (pA) (left) and $I_\text{app} = 33 + 3\eta(t)$ (pA) (right). Last row: ISI distribution of a PD-affected cell in presence of Gaussian white noise $I_\text{app} = 23 + \eta(t)$ (pA) (left) and $I_\text{app} = 23 + 3\eta(t)$ (pA) (right).  
		}\label{fig:0-8-4}	
	\end{figure}

	The main numerical results of our analysis are shown in Figures \ref{fig:0-1}--\ref{fig:0-8-4}, where  we have plotted the time evolution of the membrane potential calculated based on model \eqref{eq:1}--\eqref{eq:13}, along with the spike count profile and the corresponding spike irregularity profile. We investigate the effects of additive type of random input currents in presence of a random refractory period in a modified HH neuron under synaptic conductance dynamics. We observe that with a Poissonian spike input, the random external currents and random refractory period influence the spiking activity of a neuron in the cell membrane potential. In what follows, we use the excitatory and inhibitory conductances provided in Figure \ref{fig:0-0} for all of our simulations. 

	\clearpage
	In order to switch from healthy conditions to Parkinsonian conditions in the basal ganglia model, we consider a decrease in the current $I_\text{app}$ applied to the STN. In particular, we have $I_\text{app} = 33$ (pA) for a healthy STN cell and $I_\text{app} = 23$ (pA) for a Parkinsonian STN cell (see, e.g., \cite{So2012}). Therefore, a STN cell in the case of injected current input $I_\text{app} = 33$ (pA) results in a healthy STN cell, while a STN cell in the case of injected current $I_\text{app} = 23$ (pA) is considered as a PD-affected STN cell.
	
	\clearpage
	In Figure \ref{fig:0-1}, we have plotted the Gaussian white noise current profile, the time evolution of the membrane potential $V(t)$ with various input values of currents and refractory period for the case of a healthy STN cell ($I_\text{app} = 33$ (pA)), see e.g. in \cite{So2012}. In the second row of Figure \ref{fig:0-1}, we plot the time evolution of the membrane potential under an additive type of random input current $I_\text{app} = 33 + \eta(t)$ (pA) together with a random refractory period $t_\text{ref} = 3 + \sigma_{\text{ref}}\tilde{ \eta}(t)$ (ms). As expected, we obsevere that there are fluctuations in the time evolution of the membrane potential.  Note that a miss state occurs when a neuron is failed to spike, whereas when the state of a neuron spikes more than once within 25 (ms) we observe a burst (see, e.g., \cite{So2012}). In the last three rows of Figure \ref{fig:0-1}, there exist missing moods in the behavior of the membrane potential (e.g., from time equal to 300 to 500 (ms)). 
	This is caused by the presence of the additive noise input current and the random refractory period in the system. This is visible also in the last two rows of Figure \ref{fig:0-1}, but the fluctuations are smaller than the case shown in the second row of the same figure. Further analysis of the last two rows of Figure \ref{fig:0-1} in the case of direct input currents reveals that the time evolution of the membrane potential looks similar in both cases: direct and random refractory periods. 
	
	In Figure \ref{fig:0-2}, we switch from the healthy condition to the Parkinsonian condition by decreasing the value of $I_\text{app}$ compared to the previous cases  in Figure \ref{fig:0-1}. In particular, in the second row of Figure \ref{fig:0-2}, we consider the time evolution of membrane potential $V(t)$ with additive noise current $I_\text{app} = 23 + \eta(t)$ (pA) and random refractory period $t_\text{ref} = 3 + \tilde{ \eta}(t)$ (ms). We observe that there is an increase in the missing moods in the time evolution of the membrane potential in all last three rows. Moreover, there still exist fluctuations in the second row of Figure \ref{fig:0-2} because of the presence of additive noise input current and a random refractory period in the system. 
	In order to reduce the misses in the cases presented in Figure \ref{fig:0-2}, we applied intracellularly a DBS frequency input current $I_{\text{dbs}} = 5+5\sin(2\pi t)$ (pA) to the STN cell, as shown in Figure \ref{fig:0-3}. It is clear that the spiking activities increase in the last three rows of Figure \ref{fig:0-3}. The cases presented in Figure \ref{fig:0-3} look similar to the cases of the healthy STN cell presented in Figure \ref{fig:0-2}. There are fluctuations in the last three rows of Figure \ref{fig:0-3} due to the presence of random factors. 
	
	Next, we increase the value of $ \sigma_{\text{ref}}$ from 0.5 to 2 for the cases presented in Figures \ref{fig:0-4}--\ref{fig:0-6}. Specifically, in the presence of a random refractory period with $ \sigma_{\text{ref}}=2$, the misses and bursts are slightly increased in the second and third rows of Figure \ref{fig:0-4}, while the time evolutions in the last two rows look more stable. In the presence of a random refractory period together with a large enough value of $ \sigma_{\text{ref}}$, the healthy conditions may be switched to the Parkinsonian conditions even without decreasing the value of $I_\text{app}$. This is due to the fact that the presence of random factors could contribute to the changes in neuron responses in a cell membrane potential. 
	
	In Figure \ref{fig:0-5}, we consider similar quantities as in the cases presented in Figure \ref{fig:0-4}. The only difference is that we decrease the values of $I_\text{app}$. We observe that the silence moods seem to be  increased more than in the cases presented in the second and third rows of  Figure \ref{fig:0-4}.
	
	Similarly, we add a DBS frequency input current to the system to reduce the misses in the STN cell. In particular, in Figure \ref{fig:0-6}, we see that the presence of the DBS frequency input current leads to a significant increase in the spiking activity of the STN neuron. The cases presented in Figure \ref{fig:0-6} looks similar to the cases with healthy STN cells presented in Figure \ref{fig:0-4}. 
	
	Since the plots of Figures \ref{fig:0-4}--\ref{fig:0-6} show high firing rates, we also provide the corresponding plots to zoom these figures in Figures \ref{fig:0-6-1}--\ref{fig:0-6-3} below. When we zoom Figures \ref{fig:0-4}--\ref{fig:0-6} in the time interval $[50,500]$ (ms), we observe that the random input current and random refractory period could cause not only silences but also bursts. This is visible in the first to rows of Figures \ref{fig:0-6-1}--\ref{fig:0-6-3}. 
	
	Further analysis for investigating the spiking activities is provided based on Figures \ref{fig:0-7}--\ref{fig:0-8-4}, where we present the input-output transfer function and spike irregularity profiles of the neuron. 
	
	The phenomena observed in Figures \ref{fig:0-4}--\ref{fig:0-6} are visible also in Figure \ref{fig:0-7}. We have plotted the input-output transfer function of the neuron with the output firing as a function of average injected current in cases presented in Figures \ref{fig:0-4}--\ref{fig:0-6}. We consider the values of average injected current belonging to the interval $I_\text{average} = [22,34]$ (pA). This $I_\text{average}$ quantity includes the values of healthy STN and PD-affected STN cells. We aim at revealing the dynamic behavior of the input-output transfer function of the neuron changing over a short interval including values switched from healthy to PD conditions. However, we look mainly at the values at $I_\text{app} =23$ (pA) and $I_\text{app} = 33$ (pA). In our consideration, we first determine a set of current injection values to use as $I_\text{average} = [22,34]$ (pA). Then, for each injection level, we count the number of spikes in milliseconds to determine the firing rate. 
	Specifically, in the top left panel of Figure \ref{fig:0-7}, by considering the direct refractory period we see that the presence of additive noise input current in the system increases the spiking activity of the STN neuron more than in the case of direct input current. Looking at the top right panel of Figure \ref{fig:0-7}, it is clear that the presence of a random refractory period strongly affects the spiking activity in our system. There are fluctuations in both cases of direct input current and additive noise input current. Moreover, in the presence of the random refractory period together with the random input current, the spiking activity of the neuron increases dramatically compared with the case presented in the top left panel of Figure \ref{fig:0-7}. In the bottom left panel of Figure \ref{fig:0-7}, we investigate also the case of injecting $I_{\text{dbs}}$ current into the system. There is a significant increase in the spiking activity of the STN neuron. Furthermore, the spike count profiles in the case of direct input current look quite similar to those in both the top left and right panels of Figure \ref{fig:0-7}.  In the bottom right panel of Figure \ref{fig:0-7}, at $I_{\text{app}} = 23$ (pA) and $I_{\text{app}} = 33$ (pA) with a DBS frequency input current, the firing activity is more efficient compared to the case presented in the bottom right panel of the same figure. In the presence of DBS frequency input current, we also observe that the contribution of the random refractory period makes the spiking activity of the STN neuron more efficient compared to the case with a direct refractory period. 
	
	In Figures \ref{fig:0-8-0}--\ref{fig:0-8-4}, we look at the corresponding spike irregularity profile of the spike count in Figures \ref{fig:0-4}--\ref{fig:0-6}. In general, the variability of the ISI is measured by its coefficient of variation  $CV_\text{ISI}$. Our representative examples focus on the values at $I_\text{app} =23$ (pA) and $I_\text{app} = 33$ (pA). We first determine a set of current injection values to use as $I_\text{average} = [22,34]$ (pA). Next, the ISI is calculated by the following steps: calculate the spike times, and take the differences between spike times. Then, the $CV_\text{ISI}$ is defined as in \eqref{CV}. In particular, in the top left panel of Figure \ref{fig:0-8-0}, with direct input current and direct refractory period, we have high irregularity values of $CV_{\text{ISI}} = 1.7$ at the injected current values $I_\text{app} = 23$ (pA) and $CV_{\text{ISI}} = 1.5$ at the injected current values $I_\text{app} = 33$ (pA). In the top right panel of Figure \ref{fig:0-8-0}, using direct input current with a random refractory period, we have values of $CV_\text{ISI}$ similar to the case presented in the top left panel of the same figure. However now, there is a slight fluctuation due to the presence of the random refractory period. We look also at the ISI distribution profiles of the two cases of healthy STN and PD-affected STN cells in the last two rows of Figure \ref{fig:0-8-0}. We produce histograms of the ISI data, binned into 20 bins running from 0 to 40. The shapes of the histograms in Figure \ref{fig:0-8-0} approximate the exponential probability density function which is the probability density function for Poisson processes. Moreover, the spiking activity in both healthy and PD-affected STN cells in the case of a direct refractory period is more stable than in the case of a random refractory period.  
	
	In Figure \ref{fig:0-8-2}, we consider the cases of random input current with different values of standard deviations and in the presence of random refractory period. In particular, in the top left panel of Figure \ref{fig:0-8-2}, we see that there exist high irregularity values of the variability of the ISI with $CV_{\text{ISI}} = 1.3$ at the injected current values $I_\text{app} = 23$ (pA) and $CV_{\text{ISI}} = 1.37$ at the injected current values $I_\text{app} = 33$ (pA). However, when we increase the value of standard deviation $\sigma_{\text{app}}$ from 1 to 3, the spike trains is more regular with $CV_{\text{ISI}} = 0.87 < 1$ (the Poisson train has $CV_{\text{ISI}} = 1$) at the injected current values $I_\text{app} = 23$ (pA) and $CV_{\text{ISI}} = 0.725$ at the injected current values $I_\text{app} = 33$ (pA). The shapes of histograms still approximate the exponential probability density function. There is a decrease of the spike irregularity coefficient $\text{CV}_{\text{ISI}}$ from 1.3 to 0.87 ($I_\text{app} = 23$ (pA)) and from 1.37 to 0.725 ($I_\text{app} = 33$ (pA)) when we increase the values of $\sigma_{\text{app}}$. 
	This is due to the fact that when we increase the standard deviation of the Gaussian white noise, at some point, the fluctuations of the random input current also increase. Hence, as the input is highly oscillating, the neuron is charged up to the spike threshold and then it is reset. This essentially gives an almost regular spiking.
	In Figure \ref{fig:0-8-1}, we examine the same quantities as in the cases presented in Figure  \ref{fig:0-8-0}. The only difference is that we add a DBS frequency input current $I_{\text{dbs}}=5+5\sin(2\pi t)$ (pA) into the system. The spike irregularity is slightly reduced in both cases: direct and random refractory periods compared with the cases presented in Figure. The shapes of the histograms in Figure \ref{fig:0-8-1} also approximate the exponential probability density function. We have an increase in the spiking activity of the neuron in the cases presented in Figure \ref{fig:0-8-1}  compared with the cases presented in Figure  \ref{fig:0-8-0}.  
	
	In Figure \ref{fig:0-8-3}, we consider the cases of direct refractory period in the presence of DBS frequency input current $I_{\text{dbs}}=5+5\sin(2\pi t)$ (pA) together with random input current. In the first row of Figure \ref{fig:0-8-3}, when we add random input currents into the system, the spike irregularity values are strongly reduced compared with the cases presented in Figure \ref{fig:0-8-0}--\ref{fig:0-8-1}. In particular, in the top left panel of Figure \ref{fig:0-8-3}, we have $CV_{\text{ISI}} = 0.625$ at the injected current value $I_\text{app} = 23$ (pA) and $CV_{\text{ISI}} = 0.51$ at the injected current value $I_\text{app} = 33$ (pA). This is caused by the presence of the DBS frequency input current which makes the spiking activity of the neuron increased. However, when we increase the value of $\sigma_{\text{app}}$ from 1 to 3, the spike trains are more regular with $CV_{\text{ISI}} = 0.48 < 0.5$ at the injected current value $I_\text{app} = 23$ (pA) and $CV_{\text{ISI}} = 0.40<0.5$ at the injected current value $I_\text{app} = 33$ (pA). Note that increased ISI regularity could result in bursting \cite{Maimon2009}. The spike trains are substantially more regular with a range $CV_{\text{ISI}} \in (0; 0.5)$, and more irregular when $CV_{\text{ISI}} > 0.5$ \cite{Stiefel2013}. Therefore, the presence of random input current with high oscilations could lead to the burst discharge. 
	

	In Figure \ref{fig:0-8-4}, we analyze at the case of random refractory period in the presence of DBS frequency input current $I_{\text{dbs}}=5+5\sin(2\pi t)$ (pA) together with random input current. In the top left panel of Figure \ref{fig:0-8-4}, we see that the spike irregularity looks similar to the top left panel of Figure \ref{fig:0-8-3}. However, when we increase the value of $\sigma_{\text{app}}$ from 1 to 3, the spike irregularity increases compared with the case presented in the top right panel of Figure \ref{fig:0-8-3}. This phenomenon is quite interesting since the presence of random refractory periods could reduce the effects of random input currents on the system in the presence of DBS frequency input current. From the histograms in Figure \ref{fig:0-8-0}--\ref{fig:0-8-4} we see that their shapes approximate the exponential probability density function except the case presented in the right panel of the second row in Figure \ref{fig:0-8-4} for the healthy STN cell with a different shape. This is due to the effects of high fluctuations arising from random factors. However, the DBS treatment is commonly used in the case of Parkinsonian cells. Hence, in our consideration of DBS frequency input current, we have concentrated on the results of  Parkinsonian conditions.

	Additionally, we remark that the presence of the random refractory period and random input current affects the spiking activity of the STN neuron even in the presence of the DBS frequency input current. When we increase values of the standard deviation in random input currents and random refractory periods, the irregularity of spike trains decreases. This effect may lead to an improvement in the carrying out the information about stimulating activities in the neuron (see also, e.g., \cite{Bauermann2019}). Furthermore, the presence of random refractory periods could reduce the effects of random input currents on the system in presence of DBS frequency input current. The interplay between random refractory period and random input current in the STN cell membrane potential would contribute to further progress and model developments for the DBS therapy.

	\section{Conclusions}
	
	We have proposed a new modified HH model and described the process of synaptic conductance with random inputs. Using the description based on Langevin stochastic dynamics in a numerical setting, we analyzed the effects of random inputs in an STN cell membrane potential. Specifically, we provided details of the corresponding models along with representative numerical examples and discussed the effects of random inputs on the time evolution of the cell membrane potentials, the associated spiking activities of neurons and the spike time irregularity profiles. Our numerical results have shown that the random inputs strongly affect the spiking activities of neurons in the STN even in the presence of DBS in the system. Furthermore, we have shown that an increase in the standard deviations in the random input current can lead to a decreased irregularity of spike trains of the output neuron. However, the presence of a random refractory period together with the random input current can increase the irregularity of spike trains of the output neuron.
	More efficient managing of random factors in STN cell membrane potential models would allow for further improvements of smart treatments and bioengineering technique modalities for PD. A better understanding of cell membrane potential models developed for the targeted area of the brain that is responsible for the movement symptoms caused by PD would allow for supporting and improving the DBS therapy and other applications in the fields of biomedicine.

	\section*{Acknowledgment}
	Authors are grateful to the NSERC and the CRC Program for their
	support. RM is also acknowledging support of the BERC 2022-2025 program and Spanish Ministry of Science, Innovation and Universities through the Agencia Estatal de Investigacion (AEI) BCAM Severo Ochoa excellence accreditation SEV-2017-0718 and the Basque Government fund AI in BCAM EXP. 2019/00432.
	
	\section*{Conflict of interest}
	
	The Authors declare that there is no conflict of interest.

	%
	\section*{Supplementary (if necessary)}
	
\end{document}